\documentclass[journal,10pt]{IEEEtran}

\usepackage[T1]{fontenc}
\usepackage{lmodern}
\usepackage{microtype}
\usepackage{amsmath,amssymb,mathtools,bm}
\usepackage{amsthm}
\usepackage{booktabs,tabularx,array,multirow}
\usepackage{graphicx}
\usepackage{siunitx}
\usepackage{cite}
\usepackage{algorithm}
\usepackage{algpseudocode}
\usepackage{tikz}
\usetikzlibrary{arrows.meta,positioning,fit,calc,shapes.geometric,backgrounds,patterns}
\usepackage{pgfplots}
\usepgfplotslibrary{groupplots,fillbetween}
\pgfplotsset{compat=1.18}
\definecolor{levelone}{RGB}{31,119,180}
\definecolor{leveltwo}{RGB}{214,39,40}
\definecolor{levelthree}{RGB}{44,160,44}
\definecolor{levelfour}{RGB}{148,103,189}
\usepackage{stfloats}
\usepackage{placeins}
\usepackage{url}
\usepackage[hidelinks]{hyperref}

\usepackage[nameinlink,noabbrev]{cleveref}

\theoremstyle{definition}

\newcommand{\F}{\mathbb{F}_2}
\newcommand{\wt}{\operatorname{wt}}
\newcommand{\rank}{\operatorname{rank}}
\newcommand{\row}{\operatorname{row}}

\newcommand{\cC}{\mathcal{C}}
\newcommand{\cV}{\mathcal{V}}
\newcommand{\cE}{\mathcal{E}}
\newcommand{\cN}{\mathcal{N}}

\newcommand{\cS}{\mathcal{S}}

\newcommand{\Prb}{\mathbb{P}}

\newcommand{\xor}{\mathbin{\oplus}}

\newif\ifexpand
\expandfalse

\DeclareMathOperator*{\argmin}{arg\,min}

\title{Chromatic Block-Gibbs p-Bit Decoding of QLDPC Codes}

\author{%
\IEEEauthorblockN{%
    Mert~G\"okduman%
        \IEEEauthorrefmark{1}%
        \IEEEauthorrefmark{2}%
        \IEEEauthorrefmark{3},
    Cyril~Guyot%
        \IEEEauthorrefmark{1},
    and Henry~D.~Pfister%
        \IEEEauthorrefmark{2}%
        \IEEEauthorrefmark{3}%
        \IEEEauthorrefmark{4}%
}
\thanks{\IEEEauthorrefmark{1}Western Digital, San Jose, CA, USA.}
\thanks{\IEEEauthorrefmark{2}Duke Quantum Center, Duke University, Durham, NC, USA.}
\thanks{\IEEEauthorrefmark{3}Department of Electrical and Computer Engineering, Duke University, Durham, NC, USA.}
\thanks{\IEEEauthorrefmark{4}Department of Mathematics, Duke University, Durham, NC, USA.}
\thanks{Corresponding author: Mert G\"okduman

(e-mail: mert.gokduman@duke.edu).}}

\begin{document}
\maketitle

\begin{abstract}
We introduce a parallel decoder for quantum low-density parity-check (QLDPC) codes based on probabilistic bits (p-bits), tunable stochastic binary units controlled by local fields.  
The idea is that predicted error configurations, interpreted as explanations of the observed syndrome, are updated stochastically using local Gibbs conditionals, while a progressively strengthened parity penalty drives the search toward syndrome-valid configurations. 
A coloring of the decoding graph enables parallel block updates, and independent replicas and restarts provide trajectory diversity. 
The best candidate encountered during the search is returned, mirroring the candidate-selection strategy of recent ensemble decoders. 
At the stated update budgets, the resulting logical error rates are on par with leading decoders under both code-capacity and circuit-level noise. 
While the total number of variable resamplings can be substantially larger than the message-update count of BP-based alternatives, the numerical results establish constraint-annealed p-bit dynamics as a competitive stochastic alternative for QLDPC decoding and motivate further study of hardware-native implementations.
\end{abstract}

\vspace{-12pt}
\section{Introduction}
Quantum computing promises solutions to problems that are infeasible to solve with classical computing. 
The successful implementation of large-scale quantum algorithms requires efficient quantum error correction (QEC), as quantum information is fragile to noise~\cite{Shor1995,CalderbankShor1996,Steane1996,Gottesman1997}.
Quantum low-density parity-check (QLDPC) codes provide a route to fault-tolerant quantum computing with finite-rate encoding, sparse checks, and substantially lower asymptotic overhead than two-dimensional topological constructions \cite{BreuckmannEberhardtReview2021,TillichZemor2014,BreuckmannEberhardt2021,PanteleevKalachev2021,PanteleevKalachev2022,LeverrierZemor2022}. 
Recent finite-length constructions make this concrete by defining shallow syndrome circuits and favorable memory estimates at moderate block lengths \cite{Bravyi2024}.

The utility of QLDPC codes depends on decoders with good latency and low failure rate. Belief propagation (BP) is attractive because its cost is linear in code size and its message-passing structure is parallel and hardware-friendly~\cite{MacKay2003,MacKay2004}. 
However, QLDPC decoding differs from classical LDPC decoding in two important ways. 
First, short cycles in QLDPC decoding graphs can impair BP performance.
Second, degeneracy creates many physically distinct corrections with the same syndrome and the same logical action. 
These effects can produce oscillatory or ambiguous messages, symmetric trapping sets, and unreliable hard decisions \cite{PoulinChung2008,Roffe2020}. 

Recent QLDPC decoder work can be viewed as two complementary approaches.
One approach modifies BP so that ambiguous local reliability information is converted into a valid correction more robustly, using methods such as ordered-statistics postprocessing, stabilizer inactivation, or guided decimation
\cite{PanteleevKalachev2021,Roffe2020,HillmannLSD2025,DuCrest2022,
Yao2024BPGD,Gokduman2024ErasureBPGD, AbuLaban2023}.
A second approach deliberately introduces diversity and symmetry breaking through methods such as disordered memory effects, automorphism ensembles, multiple redundant bases, or parallel runs guided by code shortening, and then selects a valid low-cost candidate from the explored set
\cite{MuellerRelayBP2025,KoutsioumpasAutDEC2025,RabetiMBBP2026,
BhatnagarImpulse2026, Maan_2026}. 
Taken together, these methods suggest a common principle: QLDPC decoding often benefits from exploring several correction representatives and returning a low-cost valid candidate found within a finite computational budget.

Based on the same principle, we explore stochastic decoding over complete error configurations using probabilistic bits (p-bits), tunable stochastic binary units whose output bias is controlled by a local field \cite{Camsari2017, Camsari2019, KaiserDatta2021, ChowdhuryFullStack2023}. 
This connects QEC to an emerging unconventional computing framework. 
Instead of passing bit-wise beliefs about qubits or fault mechanisms, the p-bit decoder repeatedly resamples candidate corrections from local conditional distributions. 
This choice is motivated by two related observations.
First, degeneracy creates many equal or nearly equal correction representatives, so a representative-level decoder must break ambiguities during the search.
Randomized trajectory diversity is a natural way to realize many such arbitrary choices, avoiding oscillations and bit-wise belief confusion. 
Second, probabilistic computing with p-bits provides a hardware and algorithmic abstraction designed precisely for repeated biased binary sampling on sparse graphs \cite{Aadit2022}.

A p-bit is a tunable stochastic binary unit whose output is classical but randomized with a bias controlled by a local input field. 
Recently, networks of p-bits have been proposed as domain-specific probabilistic computers for sampling, optimization, machine learning, and related inference tasks \cite{Camsari2017,Camsari2019,KaiserDatta2021,Aadit2022, ChowdhuryFullStack2023,FinocchioRoadmap2024}.
Two modeling issues must be addressed to use p-bits for a QLDPC decoder.
First, the syndrome equation is a hard parity constraint over $\F$, whereas a p-bit update requires a finite real-valued input bias.
We therefore replace the hard constraint by a parity violation penalty and increase its strength during the search. 
This produces a constraint-annealed energy landscape: early updates explore low-prior-weight configurations with softened syndrome constraints, while later updates increasingly favor syndrome-valid states.
Second, unrestricted synchronous updates are not Gibbs-correct when two variables share a check. 
We resolve this by coloring the variable-conflict graph and updating only check-disjoint color classes in parallel, following the principle of chromatic Gibbs sampling \cite{Gonzalez2011}. 
This schedule aligns with the sparse-graph parallelism emphasized in p-computer architectures while preserving the exact block conditional at each fixed constraint strength.

The resulting algorithm is best viewed as a constraint-annealed stochastic-search decoder.
At a fixed constraint strength, each color-wise update is an exact Gibbs update for a softened energy. 
The complete decoder changes the constraint strength during a leg, runs multiple replicas and restarts, and records the best valid candidate encountered within a finite budget.
The p-bit decoder combines local stochastic dynamics with parallel trajectory diversity to find low-cost syndrome-valid corrections. 
Computational experiments show that this explore-and-select strategy attains logical error rates competitive with leading QLDPC decoders under both code-capacity and circuit-level noise.

The p-bit decoder is designed to utilize stochastic exploration and potentially p-bit hardware rather than to minimize the corresponding computation. 
Multiple replicas and all variables within a color class can be processed in parallel, but the large numbers of sweeps and restart legs allowed in the present experiments can require substantially more total variable updates than BP-based alternatives. 
The comparisons provided in Section \ref{sec:numerical_results} therefore reflect decoder failure rates at explicitly stated update budgets; they do not yet establish an advantage in latency, throughput, energy, or total computational work.

\section{Preliminaries}\label{sec:prelim}

\subsection{Local Gibbs conditionals and p-Bits}

The interest in p-bits is part of a broader shift toward domain-specific and unconventional computing. 
As conventional transistor scaling slows and the energy costs of sampling, optimization, and inference continue to grow, an alternative to general-purpose digital scaling is to build hardware whose native dynamics matches the structure of the target algorithms~\cite{FinocchioRoadmap2024}. 
Probabilistic computing with p-bits is one such approach that uses tunable stochastic binary units as classical random variables that can be coupled through local fields. 
A detailed full-stack overview of p-bit devices, architectures, and algorithms is given in \cite{ChowdhuryFullStack2023}.

A p-bit is a stochastic binary unit whose output is controlled by a real-valued input field. 
It may be implemented in several physical substrates, including stochastic magnetic tunnel junctions, low-barrier magnets, mixed-signal circuits, or digital stochastic logic~\cite{Raimondo2025, Borders2019, Camsari2019,Pervaiz2019Weighted,ChowdhuryFullStack2023}. 
In this paper, however, we use the p-bit at the abstraction level of a tunable Bernoulli variable. 
This abstraction is sufficient for the decoder as the hardware-specific device physics determines how the random bit and its bias are realized, while the algorithm only requires samples from the prescribed local conditional law.

In the spin convention $m_i\in\{-1,+1\}$, an ideal p-bit update has the form
\vspace{-2pt}
\begin{equation}\label{eq:pbit-spin}
    m_i
    \leftarrow
    \operatorname{sgn}\!\left[
        \tanh(I_i)-u_i
    \right],
    \qquad
    u_i\sim\operatorname{Uniform}(-1,1),
\end{equation}
which gives
\vspace{-2pt}
\begin{equation}\label{eq:pbit-prob}
    \Prb(m_i=+1\mid I_i)
    =
    \frac{1+\tanh(I_i)}{2}.
\end{equation}
Equivalently, with
\begin{equation}
    x_i=\frac{m_i+1}{2}\in\{0,1\},
\end{equation}
the p-bit produces a Bernoulli sample whose probability is set by the input field.

Let $F(x)$ be a dimensionless energy on binary variables $x\in\{0,1\}^n$, with any inverse-temperature factor absorbed into $F$. 
Holding all variables except $x_i$ fixed, define the forced-state energy difference
\begin{equation}\label{eq:forced-delta-general}
    \Delta_{10,i}
    =
    F(x_i=1,x_{\setminus i}) -
    F(x_i=0,x_{\setminus i}).
\end{equation}
The corresponding Gibbs conditional is
\begin{equation}\label{eq:gibbs-general}
    \Prb(x_i=1\mid x_{\setminus i})
    =\frac{1}{1+\exp(\Delta_{10,i})}
    = \frac{1-\tanh(\Delta_{10,i}/2)}{2}.
\end{equation}
Thus \eqref{eq:gibbs-general} is implemented by a p-bit with input
\vspace{-1pt}
\begin{equation}
    I_i=-\frac{\Delta_{10,i}}{2}.
\end{equation}

For a fixed energy and a valid update schedule, repeated conditional resampling defines a Gibbs chain with stationary distribution \cite{GemanGeman1984}
\begin{equation}
    \pi(x)\propto\exp[-F(x)].
\end{equation}  

\vspace{-2pt}
The decoder developed below uses this local conditional exactly at each fixed constraint strength. 
The complete algorithm is nonstationary because the constraint penalty changes during the search, and the output is selected from the states encountered within a finite budget. 

\vspace{-5pt}

\subsection{Sparse parity graphs and chromatic blocks}

Let $H\in\F^{m\times n}$ be a sparse parity-check matrix with variable-node set $\cV=\{1,\ldots,n\}$.
The Tanner graph of $H$ has variable nodes $j\in\cV$, check nodes $a\in\{1,\ldots,m\}$, and an edge $(a,j)$ whenever $H_{aj}=1$. 
For variable $j$, define
\begin{equation}
    \cN(j)
    =
    \{a:H_{aj}=1\},
    \qquad
    d_j
    =
    |\cN(j)|.
\end{equation}

\vspace{-1pt}
The Tanner graph determines which checks contribute to the local field of a p-bit, but it does not by itself specify which p-bits may be updated simultaneously. 
Variables that share a parity factor cannot be resampled independently in parallel from the same pre-update state while preserving the exact Gibbs kernel. 
In contrast, conditionally independent variables expose the sparse-graph parallelism targeted by p-computer architectures \cite{Aadit2022,ChowdhuryFullStack2023}. 
Therefore, we use the variable-conflict graph
\vspace{-1mm}
\begin{equation}
    G_{\mathrm{conf}}=(\cV,\cE),
\end{equation}
which contains one vertex per decoder variable and connects two variables whenever they participate in a common check:
\begin{equation}\label{eq:conflict}
    \{i,j\}\in\cE
    \quad\Longleftrightarrow\quad
    i\ne j
    \ \text{and}\
    \cN(i)\cap\cN(j)\ne\varnothing.
\end{equation}
A proper coloring
\begin{equation}
    \chi:\cV\rightarrow\{1,\ldots,n_{\mathrm{col}}\}
\end{equation}
partitions the variables into color classes
\begin{equation}
    B_g
    =
    \{j:\chi(j)=g\},
    \qquad
    g\in\{1,\ldots,n_{\mathrm{col}}\},
\end{equation}
such that no parity factor contains two variables from the same class.

For an energy composed of unary terms and row-local parity factors, this check disjointness implies conditional factorization within each color. For any $B\subseteq\cV$, let $x_B=(x_j)_{j\in B}$ denote the subvector indexed by $B$:
\begin{equation}\label{eq:block-factorization}
    \Prb\!\left(
        x_{B_g}
        \mid
        x_{\cV\setminus B_g}
    \right)
    =
    \prod_{j\in B_g}
    \Prb\!\left(
        x_j
        \mid
        x_{\cV\setminus B_g}
    \right).
\end{equation}

All p-bits in one color may therefore be resampled simultaneously, while the color classes are processed sequentially. 
This is an exact chromatic block-Gibbs update at fixed energy~\cite{Gonzalez2011}.

\subsection{Binary linear codes and stabilizer codes}

All vectors and matrix products in the decoding equations are over the binary field $\F$.
For $x\in\F^n$, $\wt(x)=\sum_j x_j$ denotes Hamming weight, and $x\xor y$ denotes bitwise addition modulo two.

A binary linear code $\cC\subseteq\F^n$ is a vector subspace. 
A parity-check matrix $H\in\F^{m\times n}$ specifies the code as
\begin{equation}
    \cC=\{x\in\F^n:Hx=0\}.
\end{equation}
If a codeword $x\in\cC$ is corrupted by an additive error $e\in\F^n$, the received vector $y=x\xor e$ has syndrome
\begin{equation}
    s=Hy=He.
\end{equation}
For a binary symmetric channel with crossover probability below one half, maximum-likelihood syndrome decoding selects a minimum-weight vector in the coset
\begin{equation}
    \{u\in\F^n:Hu=s\}.
\end{equation}
Up to a global phase, an $n$-qubit Pauli operator is represented by a binary pair $(a,b)\in\F^{2n}$ through
\begin{equation}
    D(a,b)
    =
    X^{a_1}Z^{b_1}\otimes\cdots\otimes X^{a_n}Z^{b_n}.
\end{equation}
Two such operators commute precisely when their symplectic inner product vanishes:
\begin{equation}
    \left\langle(a,b),(a',b')\right\rangle_{\mathrm{s}}
    =
    a^{\mathsf T}b'
    +
    b^{\mathsf T}a'
    =
    0.
\end{equation}

An $[[n,k]]$ stabilizer code is the joint $+1$ eigenspace of an abelian subgroup of the $n$-qubit Pauli group that does not contain $-I$ and has $n-k$ independent generators. 
If the binary representatives of these generators are collected as the rows of
\begin{equation}
    S=[S_x\mid S_z],
\end{equation}
their commutativity is equivalent to
\begin{equation}
    S_xS_z^{\mathsf T}+S_zS_x^{\mathsf T}=0.
\end{equation}

A Calderbank--Shor--Steane (CSS) code has separate $X$- and $Z$-type
check matrices
\begin{equation}
    H_x\in\F^{m_x\times n},
    \qquad
    H_z\in\F^{m_z\times n},
\end{equation}
which satisfy
\begin{equation}\label{eq:css-commute}
    H_xH_z^{\mathsf T}=0.
\end{equation}
When these matrices span the complete stabilizer spaces, the code encodes
\begin{equation}
    k=n-\rank(H_x)-\rank(H_z)
\end{equation}
logical qubits. 
The rows of matrices $L_x,L_z\in\F^{k\times n}$ may be chosen as independent $X$- and $Z$-type logical representatives satisfying
\begin{equation}\label{eq:logical-commutation}
    H_xL_z^{\mathsf T}=0,
    \qquad
    H_zL_x^{\mathsf T}=0,
    \qquad
    L_xL_z^{\mathsf T}=I_k.
\end{equation}
The first two conditions express logical--stabilizer commutation, while the third gives a canonical logical symplectic pairing~\cite{CalderbankShor1996,Steane1996,Gottesman1997}.

\subsection{Syndrome decoding model}

For the remainder of the paper, let
\begin{equation}
    H\in\F^{m\times n},
    \qquad
    A\in\F^{\nu\times n}
\end{equation}
denote a parity-check matrix and an observable matrix. 
The $n$ binary variables represent the elementary error mechanisms used by the
decoder. 
In the code-capacity setting, these variables correspond to physical-qubit errors; in a detector error model, they correspond to circuit fault mechanisms.

The measured syndrome or detector vector and the true observable vector
are
\begin{equation}\label{eq:syndrome}
    s=He,
    \qquad
    o=Ae.
\end{equation}
A decoder returns a candidate $c\in\F^n$. 
With residual
\begin{equation}\label{eq:residual}
    r=e\xor c,
\end{equation}
the candidate is valid precisely when
\begin{equation}\label{eq:validity}
    Hc=s
    \quad\Longleftrightarrow\quad
    Hr=0,
\end{equation}
and its observable prediction is correct when
\begin{equation}\label{eq:logical-residual}
    Ar=0
    \quad\Longleftrightarrow\quad
    Ac=o.
\end{equation}
Thus, exact recovery of $e$ is unnecessary: any candidate with zero residual syndrome and zero residual observable is operationally equivalent to the sampled error.

This formulation includes both numerical settings considered in this work. 
For code-capacity decoding of physical $X$ errors in a CSS code,
\begin{equation}
    H=H_z,
    \qquad
    A=L_z,
\end{equation}
and the variables represent binary $X$-error patterns on the data qubits. 
The measured syndrome is the syndrome of the $Z$-type checks, $s=H_ze$, and the observable prediction is correct whenever $L_z(e\xor c)=0$. 
The corresponding $Z$-error decoding problem is obtained by exchanging $H_z,L_z$ with $H_x,L_x$.

For circuit-level decoding, $H$ and $A$ are obtained from a detector error model (DEM). 
Each column represents one elementary fault mechanism, each row of $H$ represents a detector, and each row of $A$ represents a logical observable \cite{Derks_2025}.
The observable matrix $A$ is used to evaluate the selected candidate but does not enter the stochastic search dynamics.

\vspace{-1mm}
\subsection{Degeneracy and decoding objectives}

Two valid candidates $c$ and $c'$ are decoding-equivalent when their difference is both undetectable and logically trivial:
\begin{equation}\label{eq:equivalence}
    H(c\xor c')=0,
    \qquad
    A(c\xor c')=0.
\end{equation}
In the code-capacity $X$-error decoding problem, stabilizer-equivalent candidates satisfy
\begin{equation}
    c'=c\xor h,
    \qquad
    h\in\row(H_x),
\end{equation}
which is consistent with \eqref{eq:equivalence} by \eqref{eq:css-commute}. 
This many-to-one relation between physical errors and logical action is the relevant form of quantum degeneracy.

For independent mechanisms with $0<p_j<1/2$, define the log-likelihood ratio
\begin{equation}\label{eq:lambda}
    \lambda_j
    =
    \log\frac{1-p_j}{p_j}.
\end{equation}
Up to an additive constant, the negative log prior of a candidate is
\begin{equation}\label{eq:prior-energy}
    E_{\mathrm{prior}}(c)
    =
    \sum_{j=1}^{n}\lambda_jc_j.
\end{equation}
The most likely individual representative is therefore
\begin{equation}\label{eq:imap}
    \hat c_{\mathrm{rep}}
    =
    \argmin_{c:Hc=s}E_{\mathrm{prior}}(c).
\end{equation}
For uniform $p_j=p<1/2$, this reduces to minimum Hamming weight.

The decoder studied here uses the prior energy in \eqref{eq:prior-energy}, while its finite-budget selector retains the valid candidate of smallest selection cost encountered. 
In the reported experiments, this selection cost is Hamming weight:
\begin{equation}\label{eq:selection-cost}
    W(c)=\wt(c).
\end{equation}
For uniform priors, Hamming weight and prior energy induce the same ordering, so the selector minimizes the objective in \eqref{eq:imap} over the valid candidates encountered.
For nonuniform circuit priors, Hamming-weight selection is a representative-level heuristic rather than likelihood ranking.

\begin{table}[t]
\centering
\caption{Per-shot correction classes under the unified notation.}
\label{tab:outcomes}
\footnotesize
\renewcommand{\arraystretch}{1.18}
\begin{tabularx}{\columnwidth}{
    @{}l
    >{\raggedright\arraybackslash}p{0.31\columnwidth}
    X@{}}
\toprule
Class & Algebraic condition & Interpretation\\
\midrule
Exact
    & $r=0$
    & Exact mechanism recovery\\
Degenerate
    & $r\ne0$, $Hr=0$, $Ar=0$
    & Physically distinct but equivalent success\\
Logical
    & $Hr=0$, $Ar\ne0$
    & Valid candidate in the wrong observable class\\
Invalid-neutral
    & $Hr\ne0$, $Ar=0$
    & Invalid candidate in the correct observable class\\
Invalid-logical
    & $Hr\ne0$, $Ar\ne0$
    & Invalid candidate in the wrong observable class\\
\bottomrule
\end{tabularx}
\end{table}

\vspace{-1mm}
\section{Chromatic Block-Gibbs Decoder}\label{sec:decoder}

\subsection{Soft parity energy and constraint-strength annealing}
For a candidate vector $c\in\F^n$ and measured syndrome or detector vector $s$, define the syndrome-residual vector
\begin{equation}\label{eq:qres}
q(c)=Hc\xor s
\end{equation}
and its violation count
\begin{equation}\label{eq:violations}
V(c)=\wt(q(c))=\sum_{a=1}^{m}q_a(c).
\end{equation}
The syndrome coset is the affine set
\begin{equation}
\cS_s=\{c\in\F^n:Hc=s\}=\{c:V(c)=0\}.
\end{equation}
Rather than impose membership in $\cS_s$ through a linear solve, the decoder assigns a finite energy to parity violations:
\begin{equation}\label{eq:energy}
F_{\beta}(c;s)
=
\sum_{j=1}^{n}\lambda_jc_j
+
\beta\,\lambda_{\max} V(c),
\qquad
\lambda_{\max}=\max_j\lambda_j,
\end{equation}
where the constraint strength $\beta>0$ is varied during a search leg. 
The corresponding softened distribution is
\begin{equation}\label{eq:soft-target}
\pi_\beta(c\mid s)
=
\frac{1}{Z_\beta(s)}
\exp[-F_\beta(c;s)].
\end{equation}
At fixed $\beta$, \eqref{eq:soft-target} is a fully specified probabilistic model.
The first term is the Bernoulli negative log prior, up to an additive constant, while the second is an algorithmic relaxation of the hard parity constraint.

For uniform code-capacity noise, $p_j=p$ and $\lambda_{\max}=\lambda$, where
\begin{equation}
    \lambda=\log\frac{1-p}{p}.
\end{equation}
The normalized energy is therefore
\begin{equation}\label{eq:normalized-energy}
    \frac{F_\beta(c;s)}{\lambda}
    =
    \wt(c)+\beta V(c).
\end{equation}
Thus $\beta$ directly specifies the normalized cost of one violated constraint relative to one additional active error mechanism, independently of $p$. 
The absolute fields still scale with $\lambda$, so the p-bit updates become more deterministic at lower physical error rates.

Let $\{\beta_t\}_{t=0}^{S-1}$ be a monotone constraint-strength schedule within each restart leg. 
The code-capacity study uses
\begin{equation}\label{eq:beta-schedule}
    \beta_t
    =
    \beta_{\min}
    +
    \frac{t}{S-1}
    (\beta_{\max}-\beta_{\min}),
\end{equation}
with
\begin{equation}
    \beta_{\min}=0.175,
    \qquad
    \beta_{\max}=1.75.
\end{equation}
The circuit-level study uses the geometric schedule
\begin{equation}\label{eq:beta-schedule-geometric}
    \beta_t
    =
    \beta_{\min}
    \left(
        \frac{\beta_{\max}}{\beta_{\min}}
    \right)^{t/(S-1)}.
\end{equation}
In both cases, $\beta_t$ multiplies only the parity-violation penalty; the physical prior remains fixed. 
The process is therefore \emph{constraint annealing}, rather than temperature annealing of a fixed Hamiltonian.
\subsection{Exact one-bit p-bit field}
Consider a current bit $c_j=b\in\{0,1\}$. Let $d_j=|\cN(j)|$ be its check degree and let
\begin{equation}\label{eq:vj}
v_j=\sum_{a\in\cN(j)}q_a(c)
\end{equation}
be the number of adjacent violated checks. 
Flipping $c_j$ toggles all $d_j$ adjacent parity residuals. 
Of these, $v_j$ change from one to zero and $d_j-v_j$ change from zero to one. 
Hence
\begin{equation}\label{eq:delta-v}
\Delta V_j
=
V(c\xor \mathbf{u}_j)-V(c)
=
d_j-2v_j,
\end{equation}
where $\mathbf{u}_j\in\F^n$ is the $j$th standard basis vector.
The prior-energy change is
\begin{equation}
\Delta E_{\mathrm{prior},j}=(1-2b)\lambda_j.
\end{equation}
The complete energy change for toggling the current bit is therefore
\begin{equation}\label{eq:flip-delta}
\Delta_{\mathrm{flip},j}
=
(1-2b)\lambda_j
+
\beta\lambda_{\max}(d_j-2v_j).
\end{equation}
The forced-state difference used by a Gibbs resampling is
\begin{equation}\label{eq:delta10}
\begin{aligned}
\Delta_{10,j}
&=F_\beta(c_j=1,c_{\setminus j};s)\\
&\quad-F_\beta(c_j=0,c_{\setminus j};s)\\
&=(1-2b)\Delta_{\mathrm{flip},j}.
\end{aligned}
\end{equation}
Substituting \eqref{eq:delta10} into \eqref{eq:gibbs-general} gives the p-bit rule
\begin{equation}\label{eq:decoder-pbit}
\begin{aligned}
\Prb(c'_j=1\mid c_{\setminus j},s,\beta)
&=\frac{1}{1+\exp(\Delta_{10,j})}\\
&=\frac{1-\tanh(\Delta_{10,j}/2)}{2}.
\end{aligned}
\end{equation}
A fresh uniform random variate is compared with \eqref{eq:decoder-pbit}; the bit is set to one or zero accordingly. 
Hence, equation \eqref{eq:flip-delta} exposes the decoder's local mechanism. 
The prior biases each bit toward zero for $p_j<1/2$. 
The parity field depends only on the imbalance between satisfied and violated neighboring checks. 

\subsection{Color-wise update and finite-budget selection}

During a color update, all variables $j\in B_g$ compute their fields from the same pre-update residual $q(c)$ and are resampled independently using \eqref{eq:decoder-pbit}. 
The sampled values are then committed simultaneously. 
If a sampled bit changes, each adjacent residual bit is toggled:
\begin{equation}
    q_a\leftarrow q_a\xor 1,
    \qquad a\in\cN(j). 
\end{equation}

Because no two variables in the same color share a check, these residual updates are conflict-free within a color. 
A complete chromatic sweep processes the colors sequentially and updates every variable once.

The finite-budget decoder runs $R$ replicas over at most $L_{\max}$ restart legs, each with at most $S$ chromatic sweeps.
Replicas are initialized independently at each restart.
After initialization and after every completed sweep, the decoder inspects all replicas, adds the number of valid candidates to a running count $M$, and updates the best valid and fallback candidates.
The best valid candidate minimizes $W(c)$; the fallback minimizes $(V(c),E_{\mathrm{prior}}(c))$ over the inspected states.
The count and candidate records persist across restart legs.
The decoder stops as soon as $M\ge K$; otherwise, it runs to the budget limit and returns the best valid candidate available, or the fallback if none is valid.
All reported experiments use $K=1$.
\Cref{alg:pbit} summarizes the search; each chromatic sweep uses \eqref{eq:decoder-pbit} and the color-wise residual updates described above.

\begin{algorithm}[t]
\caption{Constraint-annealed chromatic block-Gibbs p-bit decoder}
\label{alg:pbit}
\begin{algorithmic}[1]
\Statex \textbf{Input:} $H$, syndrome $s$, priors $\{\lambda_j\}$, color classes $\{B_g\}$, initialization law $\mu_0$
\Statex \textbf{Parameters:} replicas $R$, sweeps per leg $S$, legs $L_{\max}$, valid-candidate target $K$, schedule $\{\beta_t\}$
\Statex \textbf{Output:} selected candidate $\hat c$
\State Set $M\gets0$ and clear candidate records
\For{$\ell=1,\ldots,L_{\max}$}
    \State Draw $c^{(r)}\sim\mu_0$ independently and set
    $q^{(r)}\gets Hc^{(r)}\xor s$ for $r=1,\ldots,R$
    \State Inspect all replicas; update $M$ and candidate records
    \If{$M\ge K$}
        \State \Return best valid candidate
    \EndIf
    \For{$t=0,\ldots,S-1$}
        \State Perform one chromatic sweep at strength $\beta_t$ on all $R$ replicas
        \State Inspect all replicas; update $M$ and candidate records
        \If{$M\ge K$}
            \State \Return best valid candidate
        \EndIf
    \EndFor
\EndFor
\If{$M>0$}
    \State \Return best valid candidate
\Else
    \State \Return best fallback candidate
\EndIf
\end{algorithmic}
\end{algorithm}

\subsection{Computational work, memory, and parallel depth}
\label{sec:decoder_complexity}

Let $N_H=\operatorname{nnz}(H)$ be the number of Tanner-graph edges.
Resampling variable $j$ requires $O(1+d_j)$ work for its local field, random draw, and residual update.
Counting scalar probability evaluations and random draws as constant-cost operations, one sweep therefore requires $O(n+N_H)$ work per replica. 
The coloring is computed once for a fixed $H$ and reused across syndromes.

Let $T$ be the total number of sweeps executed per replica over $\ell_{\mathrm{run}}$ restart legs, with $T\le\ell_{\mathrm{run}}S\le L_{\max}S$.
All $R$ replicas are updated in each executed sweep, giving
\begin{equation}\label{eq:number-resamplings}
    N_{\mathrm{resamp}}=RTn
\end{equation}
one-bit resamplings. Excluding initialization, the online search work is
\begin{equation}\label{eq:decoder-work}
    \mathcal{W}_{\mathrm{dec}}
    =
    O\!\left(
        RT(n+N_H)
    \right).
\end{equation}
Violation counts and selection costs can be maintained incrementally. 
Storing the shared sparse graph, priors, and coloring, the replica states and residuals, and only the best valid and fallback candidates requires
\begin{equation}\label{eq:decoder-memory}
    \mathcal{M}_{\mathrm{dec}}
    =O\!\left(N_H+R(n+m)\right)
\end{equation}
storage words.

Variables within a color and different replicas can be updated in parallel, while the colors and sweeps are processed sequentially. 
The update phase therefore has
\begin{equation}\label{eq:decoder-color-depth}
    D_{\mathrm{upd}}=Tn_{\mathrm{col}}
\end{equation}
ordered color stages. 
This stage count describes the scheduled parallelism, not end-to-end hardware latency.
For bounded row and column degrees, a greedy coloring uses $O(1)$ colors. 
If also $m=O(n)$, fixed values of $R$, $S$, and $L_{\max}$ give linear search work in $n$.
Here $n$ counts decoder variables: physical-qubit error variables for the code-capacity problem and fault mechanisms for a DEM.

\section{Numerical Results}\label{sec:numerical_results}

We evaluate the decoder under two standard noise models: independent bit-flip noise at the code-capacity level, and uniform circuit-level noise compiled to a detector error model. 
In both settings the stochastic dynamics use the operating point of \cref{sec:decoder}: chromatic block-Gibbs resampling and minimum-weight selection over the valid states collected by the replicas and restart legs. 
Each data point is accumulated until a fixed number of logical failures is observed; error bars denote one standard error.

Logical failures are counted when the predicted observable vector differs from the sampled one, $A\hat c\ne o$. In the classification of \cref{tab:outcomes}, this includes the logical and invalid-logical outcomes, but not invalid-neutral outcomes.

\begin{figure}[t]
  \centering
  \makebox[\columnwidth][c]{%
    \hspace*{-0.1\columnwidth}
    \vspace*{-1\columnwidth} \definecolor{b1BPOSD}{RGB}{31,119,180}
\definecolor{b1BP}{RGB}{255,127,14}
\definecolor{b1Impulse}{RGB}{214,39,40}
\definecolor{b1Pbit}{RGB}{148,103,189}
\definecolor{b1BPGD}{RGB}{140,86,75}

\begin{tikzpicture}
\begin{axis}[
  width=1.03\columnwidth,
  height=0.66\columnwidth,
  scale only axis=false,
  ymode=log,
  log basis y=10,
  title={B1 $[[882,24,18\le d\le 24]]$},
  xmin=0.048, xmax=0.102,
  ymin=8e-7, ymax=1.2,
  xtick={0.05,0.06,0.07,0.08,0.09,0.10},
  xlabel={Bit-flip rate $p$},
  ylabel={Failure rate},
  xlabel style={font=\scriptsize},
  ylabel style={font=\scriptsize},
  tick label style={font=\scriptsize},
  grid=both,
  minor grid style={gray!15},
  major grid style={gray!35},
  legend style={
    draw=none,
    fill=white,
    fill opacity=0.8,
    text opacity=1,
    font=\scriptsize,
    at={(0.97,0.03)},
    anchor=south east,
  },
  legend cell align={left},
  error bars/y dir=both,
  error bars/y explicit,
  error bars/error mark options={solid,line width=0.35pt},
]

\addplot+[dotted, b1BP, mark=square*, mark size=1.5pt, line width=0.55pt, mark options={solid}]
  table[x=p, y=y] {
      p y
      0.05 0.0783
      0.06 0.136
      0.07 0.247
      0.08 0.41
      0.09 0.67
      0.10 0.883
  };
\addlegendentry{BP}

\addplot+[dashed, b1BPOSD, mark=*, mark size=1.5pt, line width=0.55pt, mark options={solid}]
  table[x=p, y=y] {
      p y
      0.05 0.000772
      0.06 0.0154
      0.07 0.104
      0.08 0.345
      0.09 0.613
      0.10 0.785
  };
\addlegendentry{BP-OSD-0}

\addplot+[solid, b1BPGD, mark=triangle*, mark size=1.5pt, line width=0.55pt, mark options={solid}]
  table[x=p, y=y] {
      p y
      0.05 0.0001
      0.06 0.0031
      0.07 0.034
      0.08 0.25
      0.09 0.55
      0.10 0.83
  };
\addlegendentry{BPGD}

\addplot+[solid, b1Impulse, mark=diamond*, mark size=1.5pt, line width=0.55pt, mark options={solid}]
  table[x=p, y=y] {
      p y
      0.05 1.4e-05
      0.06 0.00077
      0.07 0.0135
      0.08 0.11
      0.09 0.341
      0.10 0.689
  };
\addlegendentry{Impulse decoding}

\addplot+[dashed, b1Pbit, mark=pentagon*, mark size=1.5pt, line width=0.55pt, mark options={solid}]
  table[x=p, y=y, y error plus=errplus, y error minus=errminus] {
      p y errplus errminus
      0.05 3.631e-06 1.1e-06 1.1e-06
      0.06 0.0001166666667 4.409585518e-05 4.409585518e-05
      0.07 0.00265 0.0003640054945 0.0003640054945
      0.08 0.0418 0.002891366459 0.002891366459
      0.09 0.2248 0.006705221846 0.006705221846
      0.10 0.6028 0.01097998179 0.01097998179
  };
\addlegendentry{p-bit decoder}

\end{axis}
\end{tikzpicture}%
    }
  \caption{Block failure rate versus bit-flip rate $p$ for the B1 $[[882,24,18\le d\le 24]]$ code under code-capacity noise. The p-bit decoder (linear $\beta:0.175\to1.75$, $256$ replicas, up to $3000$ sweeps per leg and $24$ legs) is compared with BP, BP+OSD-0~\cite{PanteleevKalachev2021}, BPGD~\cite{Yao2024BPGD}, and impulse decoding~\cite{BhatnagarImpulse2026}.}
  \label{fig:b1-code-capacity}
\end{figure}
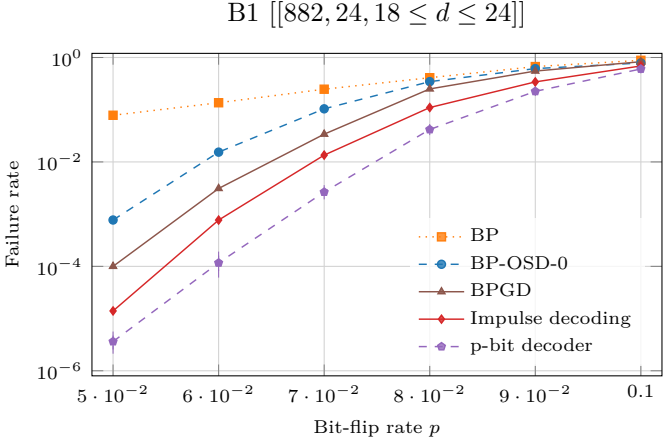

\begin{figure*}[t]
  \centering
  \input{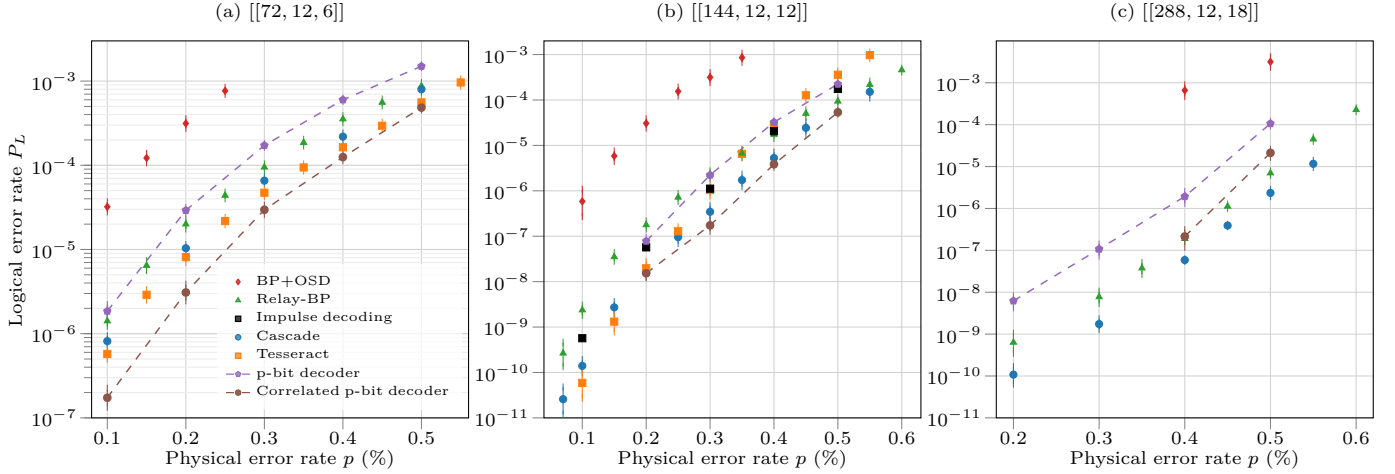}
  \vspace*{-0.05\columnwidth}
  \caption{Logical error rate per logical qubit per syndrome round versus physical error rate for the $[[72,12,6]]$, $[[144,12,12]]$, and $[[288,12,18]]$ bivariate-bicycle codes under uniform circuit-level noise. 
  The p-bit decoder acts on the DEM obtained by retaining the $X$-error component, while the correlated p-bit decoder acts on the full undecomposed DEM. Reference curves: Cascade~\cite{gu2026scalableneuraldecoderspractical}, Tesseract~\cite{beni2025tesseractsearchbaseddecoderquantum}, Relay-BP~\cite{MuellerRelayBP2025}, impulse decoding~\cite{BhatnagarImpulse2026}, and BP+OSD~\cite{PanteleevKalachev2021}.}
  \label{fig:bb-distance}
  \vspace{-12pt}
\end{figure*}

\subsection{Code-capacity noise}
\Cref{fig:b1-code-capacity} reports the block failure rate of the B1 $[[882,24,18\le d\le 24]]$ code under bit-flip noise, decoded for $X$-Pauli errors ($H=H_z$, $A=L_z$). 
We use $256$ replicas, $3000$ sweeps per leg, up to $24$ restart legs, a linear schedule $\beta:0.175\to1.75$ with candidates ranked by Hamming weight. 
Across the entire range $p\in[0.05,0.10]$ the p-bit decoder attains the lowest failure rate of all methods considered, including BP, BP+OSD-0~\cite{PanteleevKalachev2021}, belief-propagation--guided decimation (BPGD)~\cite{Yao2024BPGD}, and impulse decoding~\cite{BhatnagarImpulse2026}. 
The advantage is largest for low error rates; at $p=0.05$ the p-bit decoder reaches $3.6\times10^{-6}$, about $4\times$ below impulse decoding and more than an order of magnitude below BPGD, with the gap closing as $p$ approaches $p\approx0.09$.

\vspace{-2mm}
\subsection{Circuit benchmarks and DEM construction}\label{sec:benchmarks}

We reuse the bivariate-bicycle memory circuits of the Relay-BP paper and repository~\cite{MuellerRelayBP2025,RelayBPRepo} without modification, so the decoders are compared on identical inputs.
Each benchmark is a $d$-round memory experiment on the $[[n,12,d]]$ code generated by $A$ and $B$ over $\F[x,y]$, syndrome-extracted with a \texttt{CX} schedule and carrying $k=12$ logical observables; we use $(n,d)\in\{(72,6),(144,12),(288,18)\}$ with $A=x^3+y+y^2$, $B=y^3+x+x^2$ for $n\in\{72,144\}$ and $A=x^3+y^2+y^7$, $B=y^3+x+x^2$ for $n=288$.
The noise model is uniform circuit-level depolarizing noise at a single rate $p$: a two-qubit depolarizing channel after every \texttt{CNOT}, a single-qubit depolarizing channel on every idling qubit, a bit-flip channel in the conjugate basis after every reset, and measurement flip probability $p$.

Detector error models are built with no decomposition, so hyperedges are retained rather than split into graphlike pairs, and flattened into $H\in\F^{m\times n_{\mathrm{mech}}}$, $A\in\F^{k\times n_{\mathrm{mech}}}$ with priors $\lambda$ as in \cref{sec:decoder}. Hence, mechanisms with identical detector and observable support are merged.

For each code, we consider two circuit-level decoding problems.
The \emph{correlated} problem uses the full, undecomposed DEM, retaining the joint detector effects of circuit faults across the $X$- and $Z$-error components, including the correlations between the two components of a $Y$ fault.
The \emph{filtered} problem is obtained by applying the basis filter to the circuit before constructing its DEM. 
This retains only the $X$-error component, so the decoder does not exploit information linking the $X$ and $Z$ components of the same fault.
\begin{table}[t]
\centering
\caption{Circuit benchmarks and DEM dimensions. Every circuit runs $\text{rounds}=d$ and carries $k=12$ observables; $m$ is the detector count, $n_{\mathrm{mech}}$ the merged fault-mechanism count, and $n_{\mathrm{col}}$ the number of chromatic classes. Dimensions are independent of $p$.}
\label{tab:dems}
\footnotesize
\renewcommand{\arraystretch}{1.15}
\begin{tabular}{@{}llrrr@{}}
\toprule
Code & DEM & $m$ & $n_{\mathrm{mech}}$ & $n_{\mathrm{col}}$\\
\midrule
\multirow{2}{*}{$[[72,12,6]]$}   & filtered   & $252$  & $2\,232$   & $40$\\
                                 & correlated & $432$  & $16\,164$  & $293$\\
\midrule
\multirow{2}{*}{$[[144,12,12]]$} & filtered   & $936$  & $8\,784$   & $40$\\
                                 & correlated & $1\,728$ & $67\,824$ & $297$\\
\midrule
\multirow{2}{*}{$[[288,12,18]]$} & filtered   & $2\,736$ & $26\,208$ & $42$\\
                                 & correlated & $5\,184$ & $206\,352$ & $299$\\
\bottomrule
\end{tabular}
\vspace{-11pt}
\end{table}

\vspace{-5pt}
\subsection{Circuit-level noise}
\Cref{fig:bb-distance} reports the logical error rate per logical qubit per syndrome round for the bivariate-bicycle memory experiments. 
We use $128$ replicas, $1000$ sweeps per leg, up to $24$ restart legs, and geometric constraint-strength schedules whose endpoints are tuned per code, with $\beta_{\min}\in[0.1,0.15]$ and $\beta_{\max}\in[1.2,1.8]$.
The \emph{p-bit} and \emph{correlated p-bit} decoders act respectively on the filtered and correlated DEMs of \cref{tab:dems}, so the latter captures the cross-component fault correlations discarded by filtering.

For $[[72,12,6]]$, the correlated p-bit decoder gives the lowest logical error rate at every simulated $p$; at $p=0.1\%$ it reaches $1.7\times10^{-7}$, roughly $3.3\times$ below Tesseract and $4.7\times$ below Cascade. 
For $[[144,12,12]]$, it is on par with Cascade and stays below Relay-BP~\cite{MuellerRelayBP2025} and Tesseract through the mid-$p$ regime; for instance, at $p=0.3\%$ it attains $1.7\times10^{-7}$, about $2\times$ below Cascade and an order of magnitude below Relay-BP. 
In both cases the p-bit variant is comparable to Relay-BP but does not match the correlated p-bit decoder, indicating that access to the full correlated hyperedge model contributes materially to the gain as expected.
For the largest code $[[288,12,18]]$, the correlated p-bit decoder lowers the logical error rate by roughly $5$--$9\times$ relative to the single-component p-bit decoder, reaching $2.1\times10^{-7}$ at $p=0.4\%$; both p-bit variants nonetheless trail Cascade at this distance while remaining well below BP+OSD~\cite{PanteleevKalachev2021} (more than two orders of magnitude at $p=0.4\%$). 
Decoding the full hyperedge model at $d=18$ requires resampling a detector error model with about $2\times10^{5}$ fault mechanisms partitioned into $299$ chromatic classes.

Together, the experiments show that constraint-annealed p-bit decoding is competitive across both noise models. The correlated decoding provides substantial gains at the circuit level, although its relative performance depends on code size and operating point.

\subsection{Measured computational work}\label{sec:measured_work}

The sweep and restart limits provide a generous search allowance for our decoders, but early termination reduces the work executed on most shots. We measured the executed sweep count $T$ at $p=0.003$ using $R=128$ replicas, $S=1000$ sweeps per leg, $L_{\max}=24$ legs, and $K=1$.

Across the six configurations, the median sweep counts range from $256$ to $512$, the means from $233$ to $669$, and the empirical $99$th percentiles from $416$ to $689$.
The mean therefore uses $1.0$--$2.8\%$ of the configured $24\,000$-sweep cap.

Against the configured worst case the gap is wide: for the filtered $[[72,12,6]]$ DEM, the mean per-shot count is $6.7\times10^7$ resamplings, compared with $6.9\times10^9$ at the configured cap.

\section{Conclusion}\label{sec:conclusion}
A constraint-annealed p-bit decoder is presented for QLDPC codes.
Both code-capacity experiments and circuit-level experiments demonstrate that the method applies directly to different error models and is competitive with leading decoders on common benchmarks. 

Several questions remain before constraint-annealed p-bit decoding can be understood as a general QLDPC decoding framework. The first concerns physical realization. The proposed dynamics are compatible with stochastic binary devices, but a useful p-bit decoder requires substantially more than a source of random flips. It requires mechanisms for field computation, sparse-graph routing, enforcing a valid update schedule, and recording rare syndrome-valid states. A physical implementation should therefore be evaluated through end-to-end quantities such as energy per logically correct correction, latency tails, communication cost, and robustness to device mismatch. Hybrid architectures, in which physical p-bits provide stochastic sampling while CMOS logic implements exact parity feedback, may offer the most realistic near-term path.

A second direction is to introduce controlled disorder or asymmetry into the stochastic dynamics. The current formulation uses replicas governed by the same local objective and schedule, differing primarily through their random trajectories. State-of-the-art QLDPC decoders often obtain substantial gains by breaking symmetry more deliberately through disordered memory, perturbed priors, alternative graph representations, rewiring, or heterogeneous schedules. Analogous mechanisms could be incorporated here through replica-dependent constraint strengths, static bias perturbations, nonuniform update rates, randomized color orders, or multiple parity representations. Such asymmetry may improve logical-class coverage even when the total number of stochastic updates is held fixed.

These extensions should be accompanied by a theory of the decoder. At fixed constraint strength, the chromatic updates define a Gibbs process for a specified energy, but the annealed decoder is nonstationary and is ultimately judged by the best valid state encountered within a finite budget. Appropriate theoretical objects therefore include mixing within a fixed landscape, first-passage time to the syndrome-valid set, persistence of valid states, transitions between logical sectors, and the probability that a finite replica ensemble discovers the correct class. Establishing code- or noise-dependent conditions for rapid exploration would provide a stronger foundation than asymptotic stationary arguments alone.

Finally, the empirical scope should be extended beyond quantum-memory experiments to shallow logical circuits and universal fault-tolerant quantum computation with magic state factories.

Taken together, these directions suggest that constraint-annealed p-bit decoding is best viewed as the beginning of a broader framework for QLDPC decoding based on local stochastic updates and parallel trajectory diversity. The present results establish its viability under both code-capacity and circuit-level noise, but the framework remains at
an early stage. Resolving the open questions will determine whether constraint-annealed p-bit dynamics can mature into a general-purpose alternative to message-passing decoding.

\section{Acknowledgments}\label{sec:acknowledgments}
We thank Robert Mateescu, Jaco Hofmann, and Jack Sutton Dienhart for helpful discussions.
Any opinions, findings, conclusions, or recommendations expressed in this material are those of the authors and do not necessarily reflect the views of Western Digital.
We also acknowledge the use of generative AI tools for tasks related to programming, editing, and proofreading.
\appendices

\clearpage

\FloatBarrier
\IEEEtriggeratref{23}
\bibliographystyle{IEEEtran}
\bibliography{constraint_annealed_pbit_qldpc_long}

@article{Shor1995,
  author  = {Peter W. Shor},
  title   = {Scheme for Reducing Decoherence in Quantum Computer Memory},
  journal = {Physical Review A},
  volume  = {52},
  number  = {4},
  pages   = {R2493--R2496},
  year    = {1995},
  doi     = {10.1103/PhysRevA.52.R2493}
}

@article{CalderbankShor1996,
  author  = {A. R. Calderbank and Peter W. Shor},
  title   = {Good Quantum Error-Correcting Codes Exist},
  journal = {Physical Review A},
  volume  = {54},
  number  = {2},
  pages   = {1098--1105},
  year    = {1996},
  doi     = {10.1103/PhysRevA.54.1098}
}

@article{Steane1996,
  author  = {Andrew M. Steane},
  title   = {Error Correcting Quantum Code},
  journal = {Physical Review Letters},
  volume  = {77},
  number  = {5},
  pages   = {793--797},
  year    = {1996},
  doi     = {10.1103/PhysRevLett.77.793}
}

@phdthesis{Gottesman1997,
  author        = {Daniel Gottesman},
  title         = {Stabilizer Codes and Quantum Error Correction},
  school        = {California Institute of Technology},
  year          = {1997},
  eprint        = {quant-ph/9705052},
  archivePrefix = {arXiv}
}

@book{MacKay2003,
  author    = {David J. C. MacKay},
  title     = {Information Theory, Inference, and Learning Algorithms},
  publisher = {Cambridge University Press},
  year      = {2003}
}

@article{MacKay2004,
  author  = {David J. C. MacKay and Graeme Mitchison and Paul L. McFadden},
  title   = {Sparse-Graph Codes for Quantum Error Correction},
  journal = {IEEE Transactions on Information Theory},
  volume  = {50},
  number  = {10},
  pages   = {2315--2330},
  year    = {2004},
  doi     = {10.1109/TIT.2004.834737}
}

@article{TillichZemor2014,
  author  = {Jean-Pierre Tillich and Gilles Z{\'e}mor},
  title   = {Quantum {LDPC} Codes with Positive Rate and Minimum Distance Proportional to the Square Root of the Blocklength},
  journal = {IEEE Transactions on Information Theory},
  volume  = {60},
  number  = {2},
  pages   = {1193--1202},
  year    = {2014},
  doi     = {10.1109/TIT.2013.2292061}
}

@article{BreuckmannEberhardt2021,
  author  = {Nikolas P. Breuckmann and Jens N. Eberhardt},
  title   = {Balanced Product Quantum Codes},
  journal = {IEEE Transactions on Information Theory},
  volume  = {67},
  number  = {10},
  pages   = {6653--6674},
  year    = {2021},
  doi     = {10.1109/TIT.2021.3097347}
}

@article{BreuckmannEberhardtReview2021,
  author  = {Nikolas P. Breuckmann and Jens N. Eberhardt},
  title   = {Quantum Low-Density Parity-Check Codes},
  journal = {PRX Quantum},
  volume  = {2},
  number  = {4},
  pages   = {040101},
  year    = {2021},
  doi     = {10.1103/PRXQuantum.2.040101}
}

@article{PanteleevKalachev2021,
  author  = {Pavel Panteleev and Gleb Kalachev},
  title   = {Degenerate Quantum {LDPC} Codes with Good Finite Length Performance},
  journal = {Quantum},
  volume  = {5},
  pages   = {585},
  year    = {2021},
  doi     = {10.22331/q-2021-11-22-585}
}

@inproceedings{PanteleevKalachev2022,
  author    = {Pavel Panteleev and Gleb Kalachev},
  title     = {Asymptotically Good Quantum and Locally Testable Classical {LDPC} Codes},
  booktitle = {Proceedings of the 54th Annual ACM SIGACT Symposium on Theory of Computing},
  pages     = {375--388},
  year      = {2022},
  doi       = {10.1145/3519935.3520017}
}

@inproceedings{LeverrierZemor2022,
  author    = {Anthony Leverrier and Gilles Z{\'e}mor},
  title     = {Quantum {Tanner} Codes},
  booktitle = {2022 IEEE 63rd Annual Symposium on Foundations of Computer Science},
  pages     = {872--883},
  year      = {2022},
  doi       = {10.1109/FOCS54457.2022.00117}
}

@article{Bravyi2024,
  author  = {Sergey Bravyi and Andrew W. Cross and Jay M. Gambetta and Dmitri Maslov and Patrick Rall and Theodore J. Yoder},
  title   = {High-Threshold and Low-Overhead Fault-Tolerant Quantum Memory},
  journal = {Nature},
  volume  = {627},
  number  = {8005},
  pages   = {778--782},
  year    = {2024},
  doi     = {10.1038/s41586-024-07107-7}
}

@article{PoulinChung2008,
  author        = {David Poulin and Yeojin Chung},
  title         = {On the Iterative Decoding of Sparse Quantum Codes},
  journal       = {Quantum Information and Computation},
  volume        = {8},
  number        = {10},
  pages         = {987--1000},
  year          = {2008},
  eprint        = {0801.1241},
  archivePrefix = {arXiv}
}

@article{Roffe2020,
  author  = {Joschka Roffe and David R. White and Simon Burton and Earl T. Campbell},
  title   = {Decoding Across the Quantum Low-Density Parity-Check Code Landscape},
  journal = {Physical Review Research},
  volume  = {2},
  number  = {4},
  pages   = {043423},
  year    = {2020},
  doi     = {10.1103/PhysRevResearch.2.043423}
}

@inproceedings{DuCrest2022,
  author        = {Julien du Crest and Mehdi Mhalla and Valentin Savin},
  title         = {Stabilizer Inactivation for Message-Passing Decoding of Quantum {LDPC} Codes},
  booktitle     = {2022 IEEE Information Theory Workshop (ITW)},
  pages         = {488--493},
  year          = {2022},
  doi           = {10.1109/ITW54588.2022.9965902},
  eprint        = {2205.06125},
  archivePrefix = {arXiv},
  primaryClass  = {quant-ph}
}

@inproceedings{Yao2024BPGD,
  author        = {Hanwen Yao and {Abu Laban}, Waleed and Christian H{\"a}ger and {Graell i Amat}, Alexandre and Henry D. Pfister},
  title         = {Belief Propagation Decoding of Quantum {LDPC} Codes with Guided Decimation},
  booktitle     = {2024 IEEE International Symposium on Information Theory (ISIT)},
  pages         = {2478--2483},
  year          = {2024},
  doi           = {10.1109/ISIT57864.2024.10619083},
  eprint        = {2312.10950},
  archivePrefix = {arXiv},
  primaryClass  = {cs.IT}
}

@inproceedings{Gokduman2024ErasureBPGD,
  author        = {Mert G{\"o}kduman and Hanwen Yao and Henry D. Pfister},
  title         = {Erasure Decoding for Quantum {LDPC} Codes via Belief Propagation with Guided Decimation},
  booktitle     = {2024 60th Annual Allerton Conference on Communication, Control, and Computing},
  pages         = {1--8},
  year          = {2024},
  doi           = {10.1109/Allerton63246.2024.10735275},
  eprint        = {2411.08177},
  archivePrefix = {arXiv},
  primaryClass  = {cs.IT}
}

@article{HillmannLSD2025,
  author        = {Timo Hillmann and Lucas Berent and Armanda O. Quintavalle and Jens Eisert and Robert Wille and Joschka Roffe},
  title         = {Localized Statistics Decoding for Quantum Low-Density Parity-Check Codes},
  journal       = {Nature Communications},
  volume        = {16},
  pages         = {8214},
  year          = {2025},
  doi           = {10.1038/s41467-025-63214-7},
  eprint        = {2406.18655},
  archivePrefix = {arXiv},
  primaryClass  = {quant-ph}
}

@article{MuellerRelayBP2025,
  author        = {Tristan M{\"u}ller and Thomas Alexander and Michael E. Beverland and Markus B{\"u}hler and Blake R. Johnson and Thilo Maurer and Drew Vandeth},
  title         = {Improved Belief Propagation Is Sufficient for Real-Time Decoding of Quantum Memory},
  journal       = {arXiv preprint arXiv:2506.01779},
  year          = {2025},
  eprint        = {2506.01779},
  archivePrefix = {arXiv},
  primaryClass  = {quant-ph}
}

@article{KoutsioumpasAutDEC2025,
  author        = {Stergios Koutsioumpas and Hasan Sayginel and Mark Webster and Dan E. Browne},
  title         = {Automorphism Ensemble Decoding of Quantum {LDPC} Codes},
  journal       = {arXiv preprint arXiv:2503.01738},
  year          = {2025},
  eprint        = {2503.01738},
  archivePrefix = {arXiv},
  primaryClass  = {quant-ph}
}

@article{RabetiMBBP2026,
  author        = {Sheida Rabeti and Hessam Mahdavifar},
  title         = {Multiple-Bases Belief Propagation List Decoding for Quantum {LDPC} Codes},
  journal       = {arXiv preprint arXiv:2605.14170},
  year          = {2026},
  eprint        = {2605.14170},
  archivePrefix = {arXiv},
  primaryClass  = {cs.IT}
}

@article{BhatnagarImpulse2026,
  author        = {Shobhit Bhatnagar and Michele Pacenti and Nithin Raveendran and David Declercq and Bane Vasi{\'c}},
  title         = {Impulse Decoding of Quantum {LDPC} Codes: Equivalence of Degeneracy and Code-Shortening},
  journal       = {arXiv preprint arXiv:2606.18240},
  year          = {2026},
  eprint        = {2606.18240},
  archivePrefix = {arXiv},
  primaryClass  = {quant-ph}
}

@article{GemanGeman1984,
  author  = {Stuart Geman and Donald Geman},
  title   = {Stochastic Relaxation, {Gibbs} Distributions, and the {Bayesian} Restoration of Images},
  journal = {IEEE Transactions on Pattern Analysis and Machine Intelligence},
  volume  = {PAMI-6},
  number  = {6},
  pages   = {721--741},
  year    = {1984},
  doi     = {10.1109/TPAMI.1984.4767596}
}

@article{Camsari2017,
  author        = {Kerem Y. Camsari and Rafatul Faria and Brian M. Sutton and Supriyo Datta},
  title         = {Stochastic p-Bits for Invertible Logic},
  journal       = {Physical Review X},
  volume        = {7},
  number        = {3},
  pages         = {031014},
  year          = {2017},
  doi           = {10.1103/PhysRevX.7.031014},
  eprint        = {1610.00377},
  archivePrefix = {arXiv}
}

@article{Camsari2019,
  author  = {Kerem Y. Camsari and Brian M. Sutton and Supriyo Datta},
  title   = {p-Bits for Probabilistic Spin Logic},
  journal = {Applied Physics Reviews},
  volume  = {6},
  number  = {1},
  pages   = {011305},
  year    = {2019},
  doi     = {10.1063/1.5055860}
}

@article{Borders2019,
  author  = {William A. Borders and Ahmed Z. Pervaiz and Shunsuke Fukami and Kerem Y. Camsari and Hideo Ohno and Supriyo Datta},
  title   = {Integer Factorization Using Stochastic Magnetic Tunnel Junctions},
  journal = {Nature},
  volume  = {573},
  number  = {7774},
  pages   = {390--393},
  year    = {2019},
  doi     = {10.1038/s41586-019-1557-9}
}

@article{KaiserDatta2021,
  author  = {Jan Kaiser and Supriyo Datta},
  title   = {Probabilistic Computing with p-Bits},
  journal = {Applied Physics Letters},
  volume  = {119},
  number  = {15},
  pages   = {150503},
  year    = {2021},
  doi     = {10.1063/5.0067927}
}

@article{Aadit2022,
  author        = {Navid Anjum Aadit and Andrea Grimaldi and Mario Carpentieri and Luke Theogarajan and John M. Martinis and Giovanni Finocchio and Kerem Y. Camsari},
  title         = {Massively Parallel Probabilistic Computing with Sparse {Ising} Machines},
  journal       = {Nature Electronics},
  volume        = {5},
  number        = {7},
  pages         = {460--468},
  year          = {2022},
  doi           = {10.1038/s41928-022-00774-2},
  eprint        = {2110.02481},
  archivePrefix = {arXiv}
}

@article{ChowdhuryFullStack2023,
  author        = {Shuvro Chowdhury and Andrea Grimaldi and Navid Anjum Aadit and Shaila Niazi and Masoud Mohseni and Shun Kanai and Hideo Ohno and Shunsuke Fukami and Luke Theogarajan and Giovanni Finocchio and Supriyo Datta and Kerem Y. Camsari},
  title         = {A Full-Stack View of Probabilistic Computing With p-Bits: Devices, Architectures, and Algorithms},
  journal       = {IEEE Journal on Exploratory Solid-State Computational Devices and Circuits},
  volume        = {9},
  number        = {1},
  pages         = {1--11},
  year          = {2023},
  doi           = {10.1109/JXCDC.2023.3256981},
  eprint        = {2302.06457},
  archivePrefix = {arXiv},
  primaryClass  = {cs.ET}
}

@article{FinocchioRoadmap2024,
  author        = {Giovanni Finocchio and Jean Anne C. Incorvia and Joseph S. Friedman and Qu Yang and Anna Giordano and Julie Grollier and Hyunsoo Yang and Florin Ciubotaru and Andrii Chumak and Azad J. Naeemi and Sorin D. Cotofana and others},
  title         = {Roadmap for Unconventional Computing with Nanotechnology},
  journal       = {Nano Futures},
  volume        = {8},
  number        = {1},
  pages         = {012001},
  year          = {2024},
  doi           = {10.1088/2399-1984/ad299a},
  eprint        = {2301.06727},
  archivePrefix = {arXiv},
  primaryClass  = {cond-mat.mes-hall}
}

@article{Raimondo2025,
  author        = {Eleonora Raimondo and Esteban Garz{\'o}n and Yixin Shao and Andrea Grimaldi and Stefano Chiappini and Riccardo Tomasello and Noraica Davila-Melendez and Jordan A. Katine and Mario Carpentieri and Massimo Chiappini and Marco Lanuzza and Pedram Khalili Amiri and Giovanni Finocchio},
  title         = {High-Performance and Reliable Probabilistic {Ising} Machine Based on Simulated Quantum Annealing},
  journal       = {Physical Review X},
  volume        = {15},
  number        = {4},
  pages         = {041001},
  year          = {2025},
  doi           = {10.1103/pcmz-w776},
  eprint        = {2503.13015},
  archivePrefix = {arXiv},
  primaryClass  = {cond-mat.mes-hall}
}

@inproceedings{Gonzalez2011,
  author    = {Joseph Gonzalez and Yucheng Low and Arthur Gretton and Carlos Guestrin},
  title     = {Parallel {Gibbs} Sampling: From Colored Fields to Thin Junction Trees},
  booktitle = {Proceedings of the Fourteenth International Conference on Artificial Intelligence and Statistics},
  series    = {Proceedings of Machine Learning Research},
  volume    = {15},
  pages     = {324--332},
  publisher = {PMLR},
  year      = {2011}
}

@article{Maan_2026,
   title={Decoding correlated errors in quantum {LDPC} codes},
   volume={17},
   ISSN={2041-1723},
   url={http://dx.doi.org/10.1038/s41467-026-70556-3},
   DOI={10.1038/s41467-026-70556-3},
   number={1},
   journal={Nature Communications},
   publisher={Springer Science and Business Media LLC},
   author={Maan, Arshpreet Singh and Garcia Herrero, Francisco Miguel and Paler, Alexandru and Savin, Valentin},
   year={2026},
   month=Mar }

@article{Pervaiz2019Weighted,
  author  = {Ahmed Zeeshan Pervaiz and Brian M. Sutton
             and Lakshmi Anirudh Ghantasala and Kerem Y. Camsari},
  title   = {Weighted p-Bits for {FPGA} Implementation of
             Probabilistic Circuits},
  journal = {IEEE Transactions on Neural Networks and Learning Systems},
  volume  = {30},
  number  = {6},
  pages   = {1920--1926},
  year    = {2019},
  doi     = {10.1109/TNNLS.2018.2874565}
}

@article{Derks_2025,
   title={Designing fault-tolerant circuits using detector error models},
   volume={9},
   ISSN={2521-327X},
   url={http://dx.doi.org/10.22331/q-2025-11-06-1905},
   DOI={10.22331/q-2025-11-06-1905},
   journal={Quantum},
   publisher={Verein zur Forderung des Open Access Publizierens in den Quantenwissenschaften},
   author={Derks, Peter-Jan H.S. and Townsend-Teague, Alex and Burchards, Ansgar G. and Eisert, Jens},
   year={2025},
   month=Nov, pages={1905} }

@misc{gu2026scalableneuraldecoderspractical,
      title={Scalable Neural Decoders for Practical Fault-Tolerant Quantum Computation}, 
      author={Andi Gu and {Bonilla Ataides}, J. Pablo and Mikhail D. Lukin and Susanne F. Yelin},
      year={2026},
      eprint={2604.08358},
      archivePrefix={arXiv},
      primaryClass={quant-ph},
      url={https://arxiv.org/abs/2604.08358}, 
}

@misc{beni2025tesseractsearchbaseddecoderquantum,
      title={Tesseract: A Search-Based Decoder for Quantum Error Correction}, 
      author={Laleh Aghababaie Beni and Oscar Higgott and Noah Shutty},
      year={2025},
      eprint={2503.10988},
      archivePrefix={arXiv},
      primaryClass={quant-ph},
      url={https://arxiv.org/abs/2503.10988}, 
}

@mastersthesis{AbuLaban2023,
  author  = {{Abu Laban}, Waleed},
  title   = {Guided Decimation for Belief-Propagation Decoding of Quantum Low-Density Parity-Check Codes},
  school  = {Chalmers University of Technology},
  address = {Gothenburg, Sweden},
  year    = {2023},
  type    = {Master's thesis},
  url     = {https://odr.chalmers.se/server/api/core/bitstreams/2157505b-0ae8-4b48-b005-a8c1218ef90a/content}
}

@misc{RelayBPRepo,
  author       = {},
  title        = {{Relay-BP}: a {Rust} implementation of the {Relay-BP} decoder for {qLDPC} codes},
  year         = {2025},
  publisher    = {GitHub},
  howpublished = {\url{https://github.com/trmue/relay}},
  note         = {Accessed 2026-06-01}
}

\end{document}